\documentclass{article}

\usepackage{graphicx}
\usepackage{color}
\definecolor{red}{rgb}{1.0,0,0}
\definecolor{blue}{rgb}{0,0,1}
\usepackage{amsmath}
\usepackage[noadjust]{cite}

\begin{document}

\baselineskip 18pt

\begin{center}
{\Large {\bf  $xF_3(x,Q^2)$ Structure Function and Gross-Llewellyn Smith Sum Rule with Nuclear Effect and Higher Twist Correction}}
 \vskip5mm N. M. Nath $^{1,2}$\footnote{Corresponding author, E-mail: nmn@tezu.ernet.in},
A. Mukharjee$^{2}$, M. K. Das$^{2}$, and J. K. Sarma$^{2}$ \vskip3mm \mbox{}%
$^{1}$ Department of Physics, Rajiv Gandhi University, Rono Hills, Doimukh-791112, Arunachal pradesh, India

\mbox{}$^{2}$  High Energy Physics Laboratory, Department of Physics, Tezpur University, Tezpur-784028, Assam, India

\bigskip

\begin{abstract}
We present an analysis of the $xF_3(x,Q^2)$ structure function and Gross-Llewellyn Smith(GLS) sum rule taking into account the nuclear effects and higher twist correction. This analysis is based on the results presented in ``N. M. Nath et al., Indian J Phys 90(1), 117 (2016)". The corrections due to nuclear effects predicted in several earlier analysis are incorporated to our results of $xF_3(x,Q^2)$ structure function and GLS sum rule for free nucleon, corrected upto next-next-to-leading order (NNLO) perturbative order and calculate the nuclear structure function as well as sum rule for nuclei. In addition,  by means of a simple model we have extracted the higher twist contributions to the non-singlet structure function $xF_3(x,Q^2)$ and GLS sum rule in NNLO perturbative orders and then incorporated them to our results. Our NNLO results along with nuclear effect and higher twist corrections are observed to be compatible with corresponding experimental data and other phenomenological analysis.

\bigskip
\noindent Keywords	: Perturbative QCD; Regge theory; Nuclear effect; higher twist effect

\noindent PACS numbers: 12.39.-x ; 12.38.-t; 12.38.Bx; 13.60.Hb
\end{abstract}
\end{center}

\bigskip\noindent
{\large \bf Acknowledgements}

A. Mukharjee and M. K. Das gratefully acknowledge financial support from DAE-BRNS, India, as major research project under sanction no. 2012/ 37P/ 36/ BRNS/ 2018 dated 24 Nov 2012

\newpage
\section{Introduction}
\label{intro}
Proper understanding of the DIS(deep inelastic scattering) structure of nucleon and associated sum rules is expected to offer an important opportunity to investigate Quantum Chromodynamics(QCD) as a theory of strong interaction and hence these are regarded as the objects of intensive investigation both theoretically and experimentally in recent years(see for example \cite{NEW1} and references therein). With the recent developments of dedicated experimental facilities significant progresses have been observed in the field of experimental investigation of structure functions. Simultaneously, in this regard, tremendous progress is observed in the field of theoretical investigation with a variety of theoretical approaches.

\ Quantum Chromodynamics(QCD) is one of the most important theoretical approaches in order to account for the strong interaction processes observed at high energy particle colliders. However, the predictive power of QCD is limited. QCD is successful in describing $Q^2$ dependency of the structure functions in accord with DGLAP (Dokshitzer-Gribov-Lipatov-Altarelli-Parisi) evolution equations\cite{dglap1} in the perturbative regime i.e., within the Bjorken limit ($Q^2\gg1$, $x$ fixed and not too small). However the most important region in DIS, which has attracted much interest recently is the small-$x$ region, lies between the interface of Bjorken limit and the Regge limit.

\ The DGLAP equation is a renormalisation group equation for the quarks and gluon inside hadron. It is one of the fundamental equations of perturbative quantum chromodynamics(pQCD), being central to all theoretical predictions for lepton-hadron colliders. Solutions of DGLAP equations give the $Q^2$ evolution of both the parton distribution functions as well as various structure functions. Although QCD predicts the $Q^2$ dependence of structure functions in accord with the DGLAP equations but they have limitations on absolute prediction of structure functions. DGLAP equations cannot predict the initial values from which the evolution starts, they can only predict the evolution of structure functions with $Q^2$, once an initial distribution is given. Further, due to its complicated mathematical structure, an exact analytic determination of the structure functions is currently out of reach and one needs to apply approximated methods to arrive on predictions from the DGLAP equation. Accordingly several approximate numerical as well as semi-analytical methods for the solution of DGLAP equation have been discussed considerably over the past years (see\cite{SOL1,SOL2,SOL3,SOL4,SOL5,injp2} and references therein).

\ In our previous paper Ref. \cite{injp2}, the small-$x$ behaviour of $xF_3(x,Q^2)$ structure function and the GLS sum rule was obtained by means of solving the DGLAP evolution equation using a $Q^2$ dependent Regge behaved ansatz as initial input with pQCD corrections upto next-next-to-leading order(NNLO). In accord with  Ref.\cite{injp2} the solutions, governing the small-$x$ behaviour of $xF_3(x,Q^2)$ structure function in LO, NLO and NNLO are given by,

\begin{equation}
 F(x,t)=F(x_0,t_0)\exp\Bigg[\int_{t_0}^{t}{\Bigg(\frac{\alpha (t)}{2\pi}\Bigg)}_{LO}P(x_0,t)dt\Bigg].\Bigg(\frac{x}{x_0}\Bigg)^{(1-bt)},\label{sflo}
\end{equation}

\begin{eqnarray}
 F(x,t)=F(x_0, t_0)\exp\Bigg[\int_{t_0}^{t}{\Bigg(\frac{\alpha (t)}{2\pi}\Bigg)}_{NLO}P(x_0,t)dt \nonumber\\ +\int_{t_0}^{t}{\Bigg(\frac{\alpha (t)}{2\pi}\Bigg)}^2_{NLO}Q(x_0,t)dt\Bigg]\Bigg(\frac{x}{x_0}\Bigg)^{(1-bt)},\label{sfnlo}
\end{eqnarray}

 \noindent and

\begin{eqnarray}
 F(x,t)=F(x_0, t_0)\exp\Bigg[\int_{t_0}^{t}{\Bigg(\frac{\alpha (t)}{2\pi}\Bigg)}_{NNLO}P(x_0,t)dt \nonumber\\ +\int_{t_0}^{t}{\Bigg(\frac{\alpha (t)}{2\pi}\Bigg)}^2_{NNLO}Q(x_0,t)dt \nonumber\\+\int_{t_0}^{t}{\Bigg(\frac{\alpha (t)}{2\pi}\Bigg)}^3_{NNLO}R(x_0,t)dt\Bigg]\Bigg(\frac{x}{x_0}\Bigg)^{(1-bt)}.\label{sfnnlo}
\end{eqnarray}

\noindent respectively, where the structure function $xF_3(x,t)$ is denoted by $F(x,t)$ for nothing but simplicity, where $t=\ln⁡ \bigg(\frac{Q^2}{\Lambda^2}\bigg)$ and $\Lambda$ is the QCD cut off parameter. Here

\begin{eqnarray}
P(x,t)=\frac{2}{3}\{3+4ln(1-x)\}
+\frac{4}{3}\int_{x}^{1} \frac{d\omega}{1-\omega}\Bigg\{\frac{1+\omega^2}{\omega}\omega^{-(1-b_i^{NS} t)}-2\Bigg\},
\end{eqnarray}

\begin{eqnarray}
 Q(x,t)= \int_{x}^{1} \frac{d\omega}{\omega}P^{(1)}(\omega){\omega}^{-(1-b t)},
\end{eqnarray},

\noindent and

\begin{eqnarray}
 R(x,t)=\int_{x}^{1} \frac{d\omega}{\omega}P^{(2)}(\omega){\omega}^{-(1-b t)}
\end{eqnarray}

\noindent in which the two loop and three loop correction terms to the splitting functions for non-singlet structure functions are given by\cite{SPFUP}

\begin{eqnarray}
 P^{(1)}(\omega)= C_F^2\bigg[P_F(\omega)-P_A(\omega)+\delta(1-\omega)\{\frac{3}{8}-\frac{1}{2}\pi^2+\zeta(3)-8\tilde{S}(\infty)\} \bigg]\nonumber\\+\frac{1}{2}C_FC_A\bigg[P_G(\omega)+P_A(\omega)+\delta(1-\omega)\bigg\{\frac{17}{12}+\frac{11}{9}\pi^2-\zeta(3)+8\tilde{S}(\infty)\bigg\}\bigg]\nonumber\\+C_F T_RN_F\bigg[P_{N_F}(\omega)-\delta(1-\omega)\bigg\{\frac{1}{6}+\frac{2}{9}\pi^2\bigg\}\bigg]
\end{eqnarray}

\noindent and

\begin{eqnarray}
P^{(2)}(\omega)= N_F\bigg[-183.187D_0-173.927\delta(1-\omega)-\frac{5120}{81}L_1-197.0\nonumber\\+381.1\omega+72.94\omega^2+44.79\omega^3-1.497\omega L_0^3-56.66L_0L_1\nonumber\\-152.6L_0-\frac{2608}{81}L_0^2-\frac{64}{27}L_0^3\bigg]\nonumber\\ +N_F^2\frac{64}{81}\bigg[-D_0-\bigg(\frac{51}{16}+3\zeta_3-5\zeta_2\bigg)\delta(1-\omega)+\frac{\omega}{1-\omega}L_0\bigg(\frac{3}{2}+5\bigg)\nonumber\\+1 +(1-\omega)\bigg(6+\frac{11}{2}L_0+\frac{3}{4}L_0^2\bigg)\bigg],
 \end{eqnarray}

\noindent with

\begin{eqnarray}
P_F(\omega)=-2\frac{1+\omega^2}{1-\omega}\ln \omega \ln (1-\omega) - \bigg(\frac{3}{1-\omega}+2\omega\bigg)\ln \omega - \frac{1}{2}(1+\omega)\ln^2\omega\nonumber\\-5(1-\omega),
\end{eqnarray}

\begin{eqnarray}
P_G(\omega)=\frac{1+\omega^2}{(1-\omega)_+}\bigg[\ln^2\omega+\frac{11}{3}\ln \omega+\frac{67}{9}-\frac{1}{3}\pi^2\bigg]+2(1+\omega)\ln \omega \nonumber\\+ \frac{40}{3}(1-\omega),
\end{eqnarray}

\begin{eqnarray}
P_{N_F}(\omega)=\frac{2}{3}\bigg[\frac{1+\omega^2}{(1-\omega)_+}(-\ln \omega -\frac{5}{3})-2(1-\omega)\bigg],
\end{eqnarray}

\noindent and

\begin{eqnarray}
P_{A}(\omega)= 2\frac{1+\omega^2}{1+\omega}\int_{\omega/(1+\omega)}^{1/(1+\omega)}\frac{dz}{z}\ln \frac{1-z}{z} + 2(1+\omega)\ln \omega +4(1-\omega).
\end{eqnarray}

\noindent Here the following abbreviations are used,

\begin{eqnarray}
D_0=\frac{1}{(1-\omega)_+}, \hspace{5mm} L_1 = \ln (1-\omega), \hspace{5mm} L_0=\ln \omega.
\end{eqnarray}

\noindent The results for small-$x$ behaviour of structure function in accord with eq.(\ref{sflo}), (\ref{sfnlo}) and (\ref{sfnnlo}) were observed to be consistent with the available experimental data taken from CCFR\cite{r24,r25},  NuTeV \cite{r26}, CHORUS\cite{r27} and CDHSW\cite{r28} collaborations as well as with several other strong analysis performed by MRST98\cite{mrs98}, CTEQ4\cite{cteq4}, KPS\cite{KPS} and KS\cite{glsKS} for $xF_3(x,Q^2)$ structure functions. The phenomenological success achieved in this regard inspired us to utilise these results in determining the GLS sum rule, which is associated with $xF_3(x,Q^2)$ structure function.

\ Considering above relations for the small-$x$ behaviour of $xF_3(x,Q^2)$ structure function we obtained the GLS integral with LO, NLO and NNLO corrections as

 \begin{eqnarray}
\int_{x'}^1\frac{xF_3(x,Q^2)}{x}dx=T_{GLS}-\int_0^{x'}\frac{dx}{x}F(x_0,t_0)\exp\Bigg[\int_{t_0}^{t}{\Bigg(\frac{\alpha(t)}{2\pi}\Bigg)}_{LO}\nonumber\\P(x_0,t)dt\Bigg].
\Bigg(\frac{x}{x_0}\Bigg)^{(1-bt)},\label{srlo}
 \end{eqnarray}

 \begin{eqnarray}
\int_{x'}^1\frac{xF_3(x,Q^2)}{x}dx=T_{GLS}-\int_0^{x'}\frac{dx}{x}F(x_0, t_0)\exp\Bigg[\int_{t_0}^{t}{\Bigg(\frac{\alpha (t)}{2\pi}\Bigg)}_{NLO}P(x_0,t)dt\nonumber\\+\int_{t_0}^{t}{\Bigg(\frac{\alpha (t)}{2\pi}\Bigg)}^2_{NLO}Q(x_0,t)dt\Bigg]\Bigg(\frac{x}{x_0}\Bigg)^{(1-bt)}\label{srnlo}
 \end{eqnarray}

 \noindent and

 \begin{eqnarray}
\int_{x'}^1\frac{xF_3(x,Q^2)}{x}dx=T_{GLS}-\int_0^{x'}\frac{dx}{x}F(x_0, t_0)\exp\Bigg[\int_{t_0}^{t}{\Bigg(\frac{\alpha (t)}{2\pi}\Bigg)}_{NNLO}P(x_0,t)dt \nonumber\\+\int_{t_0}^{t}{\Bigg(\frac{\alpha (t)}{2\pi}\Bigg)}^2_{NNLO}Q(x_0,t)dt \nonumber\\+\int_{t_0}^{t}{\Bigg(\frac{\alpha (t)}{2\pi}\Bigg)}^3_{NNLO}R(x_0,t)dt\Bigg]\Bigg(\frac{x}{x_0}\Bigg)^{(1-bt)}\label{srnnlo}
 \end{eqnarray}

 \noindent respectively and they were also observed to be compatible with the CCFR\cite{r25} experimental data and KS \cite{glsKS} analysis.

\ Detailed phenomenological analysis performed in Ref. \cite{injp2} revealed that higher order pQCD corrections have a significant contribution towards the precise predictions of the structure functions as well as the sum rules. However recent analysis indicates that precise prediction of structure functions demand to incorporate several non-perturbative effects, in addition to pQCD corrections. There are several non-perturbative effects such as nuclear effects, higher twist effects, target mass corrections(TMC) etc., to be incorporated into the joint QCD analysis of DIS structure functions and sum rules. However in accord with \cite{tmc} the contribution due to TMC within the region of our consideration is neglected. In this paper we present an analysis of the NNLO results of $xF_3(x,Q^2)$ structure function and the GLS sum rules taking into account the nuclear effects and Higher twist corrections.

\section{Nuclear Shadowing Effect in $xF_3(x,Q^2)$ structure function and GLS sum rule}

\subsection{Nuclear shadowing effect in $xF_3(x,Q^2)$ structure function}

The fact that the structure functions of bound and free nucleons are not equal was discovered in a deep inelastic muon experiment carried out by the European Muon Collaboration at CERN in 1982\cite{EMC}. Since then the nuclear effect has been actively investigated with ever more sophisticated and ingenious deep inelastic scattering experiment with charged leptons and neutrinos.

\ Available experimental information on nuclear structure functions are mainly from charged-lepton DIS experiments performed at CERN, SLAC, DESY, FNAL and at JLab. In addition, data from the Drell–Yan reaction of protons off nuclear targets are also available \cite{b014}. The experiments usually measures the ratio $R_2(x,Q^2)$ of the structure function $F_2(x,Q^2)$ of a complex nucleus to deuterium.  The studies on the behaviour of the ratio $R_2(x,Q^2)$ as a function of $x$ for a given fixed $Q^2$ reflects four distinct region of characteristic nuclear effects: shadowing region($x<0.1$), anti-shadowing region($0.1 < x < 0.3$), EMC region ($0.3<x<0.8$) and fermi motion region($x>0.8$).

\ In addition to the charged-lepton DIS, neutrino DIS have also been a significant process for investigating the structures of hadrons and nuclei(see \cite{KPNPA,KPPRD,KPNPA98,QV,HKN,nxf31,nxf32} and their references). Many $\nu$-DIS experimental programmes, such as BEBC\cite{COOPER9}, CDHS\cite{CDHS10}, E545\cite{HANLON11}, and, etc.,were carried out in order to have proper information about structure of nucleon, nuclear effect(EMC), mixing angle of weak interaction, etc. But none of them were able to confirm the EMC effect due to the presence of statistical uncertainties with the measurements. However it is expected that the measurements of differential cross-sections and structure functions in CCFR\cite{r24,r25}, NuTeV\cite{r26}, CDHSW\cite{r28}, CHORUS\cite{r27} and more recently in MINERvA\cite{minerva1,minerva2}, through  $\nu$-DIS experiments would provide us better understanding about nuclear effect as well as internal structure of nucleon(see \cite{dcg} for more details).

\ Along with the experimental efforts, several groups have been performed theoretical as well as phenomenological analysis of the nuclear effects in neutrino-nucleus DIS. Among them most prominent are the Kulagin and Petti(KP)\cite{KPNPA,KPPRD,KPNPA98}, Qiu and Vitev(QV)\cite{QV} and Hirai, Komano and Naga(HKN) groups\cite{HKN,HKN07}, which have predicted the nuclear corrections in the low $x$ region. Kulagin and Petti's approach is quite different from the above ones in the sense that they try to calculate the nuclear corrections in conventional nuclear models as far as they can, and then they try to attribute remaining factors to off-shell effects of bound nucleons for explaining the data.

\ For simplicity, let us denote $xF_3(x,t)$ as $F(x,t)$ and with this notation our results, (\ref{sfnnlo}) which predicts the $xF_3(x,Q^2)$ structure functions for a nucleon(single or free) in NNLO pQCD corrections can be written as

\begin{eqnarray}
F(x,t)=F(x_0,t_0) \exp\Bigg[\int_{t_0}^{t}{\Bigg(\frac{\alpha (t)}{2\pi}\Bigg)}_{NNLO}U(x_0,t)dt \nonumber\\ +\int_{t_0}^{t}{\Bigg(\frac{\alpha (t)}{2\pi}\Bigg)}^2_{NNLO}V(x_0,t)dt \nonumber\\ +\int_{t_0}^{t}{\Bigg(\frac{\alpha (t)}{2\pi}\Bigg)}^3_{NNLO}W(x_0,t)dt\Bigg]\Bigg(\frac{x}{x_0}\Bigg)^{(1-bt)}.
\label{8ch1}
\end{eqnarray}

\noindent However in predicting the free nucleon structure functions, we need to consider the input point $F(x_0,t_0)$, a free nucleon structure function at $x=x_0$ and $t=t_0$. In our previous analysis performed in Ref.\cite{injp2}, the points were taken from the available experimental data. It is known that the experimental data for nucleon structure functions are extracted from nuclear targets and hence they are with several nuclear effects. Thus the experimental input points, we considered in our previous analysis are nothing but nuclear structure function $F^A(x_0,t_0)$, which in turn leads to inaccuracy in predicting free nucleon structure function. Therefore accurate prediction of free nucleon structure function requires a nuclear effect free input point.

\ The experimental results are the structure functions for bound nucleon $F^A(x,t)$ which is related to the free nucleon structure function as

 \begin{eqnarray}
R(x,t)=\frac{F^A(x,t)}{F^N(x,t)}.
\label{ne2}
 \end{eqnarray}

\noindent Here $F^A(x,t)$ represents the nuclear structure function per nucleon and $F^N(x,t)$, the free nucleon structure function. At $x=x_0$ and $t=t_0$, if we consider the value of the nuclear correction factor to be $R(x,t)=R_0$, the input point in (\ref{8ch1}) can be replaced with $F^N(x_0,t_0)=\frac{F^A(x_0,t_0)}{R_0}$ and provides

\begin{eqnarray}
F^N(x,t)=\frac{F^A(x_0,t_0)}{R_0} \exp\Bigg[\int_{t_0}^{t}{\Bigg(\frac{\alpha (t)}{2\pi}\Bigg)}_{NNLO}U(x_0,t)dt \nonumber\\ +\int_{t_0}^{t}{\Bigg(\frac{\alpha (t)}{2\pi}\Bigg)}^2_{NNLO}V(x_0,t)dt \nonumber\\ +\int_{t_0}^{t}{\Bigg(\frac{\alpha (t)}{2\pi}\Bigg)}^3_{NNLO}W(x_0,t)dt\Bigg]\Bigg(\frac{x}{x_0}\Bigg)^{(1-bt)}.
\label{ne3}
\end{eqnarray}

\begin{figure*}
\centering{
\includegraphics[scale=0.5]{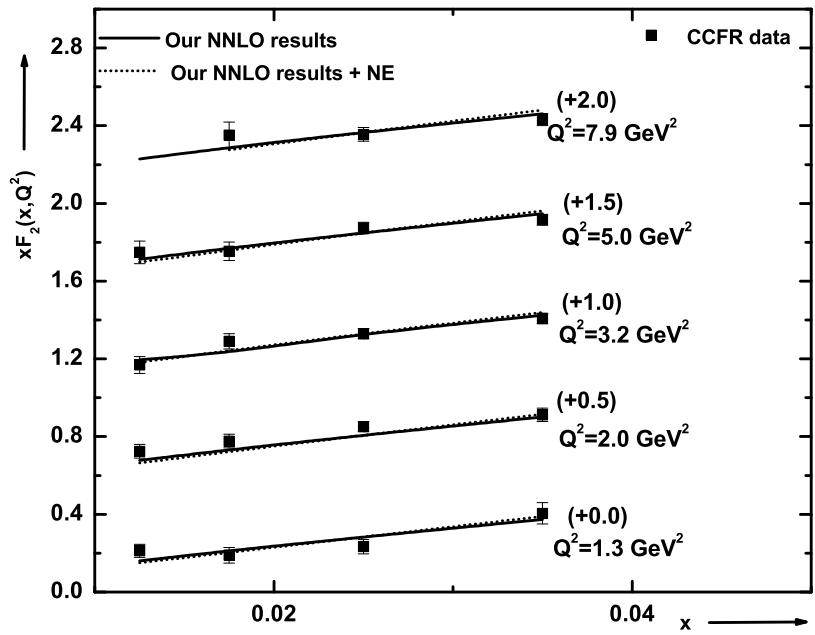}}

\caption{Our NNLO results for $xF_3(x,Q^2)$ structure function with and without nuclear effect, in comparison with the CCFR\cite{r24} data.}
\label{NEF3NS}       
\end{figure*}

\noindent Above expression is capable of predicting the free nucleon structure function through the experimental data $F^A(x_0,t_0)$ along with the correction factor $R_0$.

\ Moreover, due to the unavailability of free nucleon structure function data, direct phenomenological analysis of (\ref{ne3}) is not possible. In order to perform phenomenological analysis of our results with the experimental data either we need to remove nuclear effects from the data points or include the corresponding effects to our results of free nucleon. Here we have considered the later one, i.e., we have incorporated the nuclear correction factor $R(x)$ with our calculations as

\begin{eqnarray}
F^A(x,t)=R(x)F^N(x,t)=R(x)\frac{F^A(x_0,t_0)}{R_0} \exp\Bigg[\int_{t_0}^{t}{\Bigg(\frac{\alpha (t)}{2\pi}\Bigg)}_{NNLO}U(x_0,t)dt \nonumber\\ +\int_{t_0}^{t}{\Bigg(\frac{\alpha (t)}{2\pi}\Bigg)}^2_{NNLO}V(x_0,t)dt \nonumber\\ +\int_{t_0}^{t}{\Bigg(\frac{\alpha (t)}{2\pi}\Bigg)}^3_{NNLO}W(x_0,t)dt\Bigg]\Bigg(\frac{x}{x_0}\Bigg)^{(1-bt)},
\label{ne4}
\end{eqnarray}

\noindent in order to describe properly the experimental results.

\ In this paper we have utilised the results for the nuclear correction factor $R(x)$ predicted in KP\cite{KPNPA}. Incorporating the corresponding corrections to our calculations of $F(x,t)$ structure function, we have obtained the nuclear structure function $F^{(A)}(x,t)$ and depicted them in Fig. \ref{NEF3NS}. Here we have shown only the modification of our NNLO results in comparison with CCFR, NuTeV, CHORUS and CDHSW experimental data. We observe that our results for free nucleon structure functions, along with nuclear effect predicted by KP provides a well description of available experimental data for nuclear structure functions.

\subsection{Nuclear shadowing effect in GLS sum rule}

\begin{figure*}
\centering{
\includegraphics[scale=0.5]{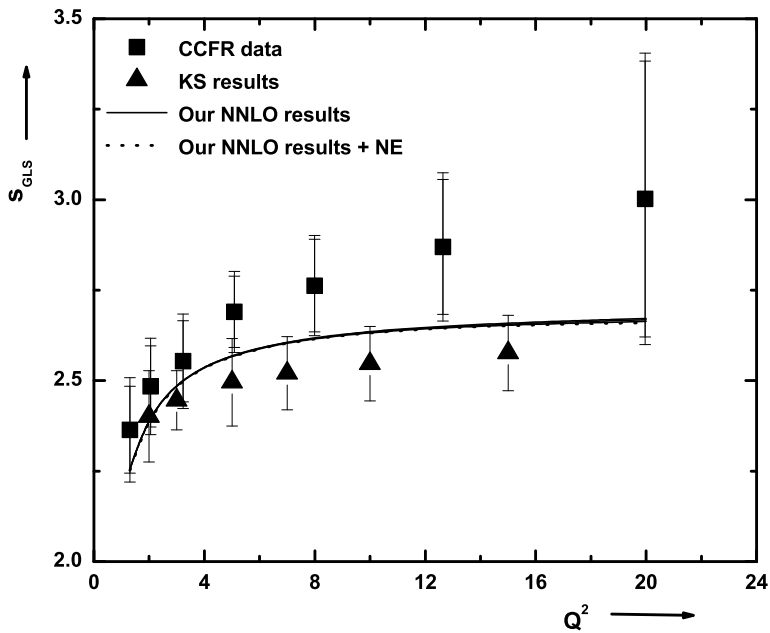}}

\caption[Our NNLO results for Gross-Llewelln Smith sum rule with and without nuclear effect, in comparison with those of CCFR measurements.]{Our NNLO results for Gross-Llewelln Smith sum rule with and without nuclear effect, in comparison with those of CCFR measurements. ($Q^2$'s are taken in the unit of $GeV^2$).}
\label{NEGLSQ}       
\end{figure*}

Analogous to the structure functions, experimental determination of the DIS sum rules consists of considerable nuclear effects. As DIS sum rules are associated with the underlying symmetry as well as conservation laws of interactions, they provide strong normalization constraints on the structure functions. Therefore the sum rules are expected to provide an important bridge between different nuclear effects. In this section we briefly discuss the nuclear effects in GLS sum rule based on several earlier analysis. We then incorporate possible nuclear corrections to our results Eq. (\ref{srnnlo}) for GLS sum rule, obtained in the previous paper\cite{injp2} and perform phenomenological analysis in comparison with the experimental measurements.

\ Experimental measurements of GLS sum rule was performed by CCFR and the results were extracted from $Fe$ target. In order to compare our results for GLS sum rule obtained in \cite{injp2}, we refer the nuclear corrections estimated in \cite{KPPRD,KPNPA98}. The detailed investigation on the nuclear corrections to GLS sum rule was performed in Ref. \cite{KPPRD}. They explicitly separated the nuclear corrections to the GLS integral as $S_{GLS}^A = S_{GLS}^N + \delta S_{GLS}$, where $S_{GLS}^N$ refers to the GLS integral for nucleon. In accord with their predictions, the nuclear corrections to the GLS sum rule cancel out as $x\rightarrow 0$ in the leading order, which is due to the baryon charge conservation. They have also calculated the GLS integral, $S_{GLS}$ for different nuclear targets. In Ref. \cite{KPPRD,KPNPA98}, they obtained the corrections for iron and deuteron nuclei as $\frac{\delta S_{GLS}^{Fe}}{3} = -\frac{4.0 \times 10^{-3}}{Q^2}$ and $\frac{\delta S_{GLS}^{D}}{3} = -\frac{6.3 \times 10^{-4}}{Q^2}$ respectively. In Ref. \cite{KPPRD} they have nicely presented their result in Fig. 10. From Fig. 10 we observe that the nuclear correction $\delta_{GLS}$ decreases progressively by increasing $Q^2$.

\ The GLS sum rule for nuclei can be expressed as

\begin{eqnarray}
S_{GLS}^A(x_{min},Q^2)\bigg|_{NNLO}=S_{GLS}^N(Q^2)+\delta S_{GLS},
 \end{eqnarray}

\noindent where the first term on the right hand side of above equation represents the GLS sum rule for free nucleon and the second term for the nuclear correction. Using the NNLO pQCD corrected expression (\ref{srnnlo}) as $S_{GLS}^N(Q^2)$ we get

\begin{eqnarray}
S_{GLS}^A(x_{min},Q^2)\bigg|_{NNLO}=S_{GLS}(Q^2)\bigg|_{NNLO}-\int_0^{x_{min}}\frac{dx}{x}\Bigg[F_3^{NS}(x_0, t_0)\Bigg(\frac{x}{x_0}\Bigg)^{(1-bt)}\nonumber\\\exp\bigg\{\int_{t_0}^{t}{ \Bigg(\frac{\alpha (t)}{2\pi}\Bigg)}_{NNLO}\nonumber\\ P(x_0,t)dt + \int_{t_0}^{t}{\Bigg(\frac{\alpha (t)}{2\pi}\Bigg)}^2_{NNLO}Q(x_0,t)dt \nonumber\\ + \int_{t_0}^{t}{\Bigg(\frac{\alpha (t)}{2\pi}\Bigg)}^3_{NNLO}R(x_0,t)dt\Bigg\}\bigg]+\delta S_{GLS}.
 \end{eqnarray}

\ Now incorporating the KP\cite{KPPRD,KPNPA98} prediction $\frac{\delta S_{GLS}^{Fe}}{3} = -\frac{4.0 \times 10^{-3}}{Q^2}$, for the nuclear correction term, we have calculated $S_{GLS}^A$ and depicted the results in Fig. \ref{NEGLSQ}, in comparison with CCFR measurements of GLS sum rule with $Fe$ as the target. In addition, we have plotted our NNLO results and the results of KS\cite{glsKS} prediction. From the figure we see that the our NNLO expression for GLSSR along with necessary nuclear correction has the capability of describing the experimental data of GLSSR for nuclei.

\section{Higher Twist Corrections on structure functions}

The behaviour of the deep inelastic structure functions can be analyzed with the perturbative QCD.  A method used for this analysis is the operator product expansion method(OPE)\cite{OPE}. The OPE is successful in describing the contributions from different quark-gluon operators to hadronic tensor and helps in ordering them according to their twist. In accord with OPE, the DIS structure functions and sum rules consist of two parts, the leading twist(LT) and the higher twist(HT) contributions:

 \begin{eqnarray}
F(x,Q^2) = F^{LT}(x,Q^2) +\frac{H_i(x,Q^2)}{Q^2},
\label{EQ:9CH1}
 \end{eqnarray}

\noindent where $i$ labels the type of the structure function ($F=F_2, F_3, g_1$). The leading twist term is associated with the single particle properties of quarks and gluons inside the nucleon and is responsible for the scaling of DIS structure function via perturbative QCD $\alpha_s(Q^2)$ corrections. The higher twist terms reflect instead the strength of multi-parton interactions ($qq$ and $qg$). Since such interactions spoil factorization one has to consider their impact on the parton distribution functions extracted in the analysis of low-$Q^2$ data.  Because of the non-perturbative origin it is difficult to quantify the magnitude and shape of the higher twist terms from first principles and current models can only provide a qualitative description for such contributions, which must then be determined phenomenologically from data.

\ The higher twist terms are governed by the terms contributing at different orders of $1/Q^2$:

 \begin{eqnarray}
\frac{H_i(x,Q^2)}{Q^2}= \frac{h_1(x)}{Q^2}+\frac{h_2(x)}{Q^4}+............,
\label{EQ:9CH2}
 \end{eqnarray}

\noindent the leading term in this expansion is known as twist-two, the sub-leading ones twist-three, etcetera. The higher twist terms are suppressed by terms of order $1/Q^2$, $1/Q^4$..., respectively.

\ The currently available experimental measurements of deep inelastic structure functions covers a wide range of $x$ and $Q^2$ with ever increasing precision, which lead to an interesting challenge for theoretical physics in describing these data in the low-$Q^2$ domain. pQCD predictions, even with higher order corrections up to NNLO and NNNLO observed to be not sufficient for a precise description of deep inelastic structure function data, which in turn reveals that the discrepancy among data and pQCD predictions are not primarily the sub-leading terms in powers of $\alpha_s$, but corrections which are proportional to the reciprocal value of the photon virtuality $Q^2$, viz. higher-twist terms\cite{jbcb}.

\ The extraction of higher twist terms from the data is a longstanding problem, as recognized from the very first developments of a pQCD phenomenology \cite{0304210ab16,rujula}. Existing information about higher twist terms in lepton-nucleon structure functions is scarce and somewhat controversial. Early analysis \cite{akp1,akp2} suggested a significant HT contribution to the longitudinal structure function $F_L$. The subsequent studies with both charged leptons \cite{akp3,akp4,akp5} and neutrinos \cite{KPS} raised the question of a possible dependence on order of QCD calculation used for the leading twist.  The common wisdom is generally that HTs only affect the region of $Q^2 \sim 1 - 3 GeV^2$ and can be neglected in the extraction of the leading twist.

\ The higher twist terms are presently poorly known and currently is a subject of both theoretical and phenomenological studies(see \cite{htnon} and the references therein). A better understanding of HT terms, in particular their role in describing low $Q^2$ and high $x$ DIS data is important and provides valuable information on quark gluon correlations inside the nucleon.  The importance of higher twist (HT) contribution to structure functions was pointed from the very beginning of QCD in comparison with experimental data\cite{0304210ab16} on structure functions. Several reports are available on the determination of the higher twist contributions in the electron-DIS structure functions $F_2^{ep,ed}(x,Q^2)$(see \cite{9607275s5,0807.0248ht9,blb} for details)as well as neutrino-DIS structure function $xF_3(x,Q^2)$ \cite{9607275s6,0807.0248ht10a,0807.0248ht10b,0807.0248ht10c,0807.0248ht10d}. Further, in general the higher twist corrections are also present in the case of polarized deeply inelastic scattering. But in this regard as the polarized structure functions are predicted in terms of an asymmetry, the effect of higher twist corrections in the denominator function needs to be known accurately. However, in \cite{0807.0248ht11} no significant higher twist contributions were found. On the other hand some authors predicted (see for example Ref.\cite{0807.0248ht12}) existence of higher twist contributions in the low $x$ region, which is also the region of very low values of $Q^2$.

\ The usual approach in analyses whose main aim is the extraction of leading twist PDFs is either to parametrize the higher twist contributions by a phenomenological form and fit the parameters to the experimental data\cite{a15,0911.2254ackmmmo2}, or to extract the $Q^2$ dependence by fitting it in individual bins in $x$ \cite{akp5,9607275s5,0807.0248ht9,0911.2254ackmmmo4,0911.2254ackmmmo6}. Such an approach effectively includes contributions from multiparton correlations (the true higher twist contributions) along with other power corrections that are not yet part of the theoretical treatment of DIS at low $Q^2$. These include $O(1/Q2)$ contributions such as jet mass corrections \cite{0911.2254ackmmmo8} and soft gluon resummation \cite{0911.2254ackmmmo9}, as well as contributions which are of higher order in $\alpha_s$ but whose logarithmic $Q^2$ behavior mimics terms $\propto \frac{1}{Q^2}$ at low virtuality\cite{0911.2254ackmmmo10,0807.0248ht9}.

\ In the following subsections we present a simple model in order to extract the higher twist contribution to $xF_3(x,Q^2)$ structure function and GLS sum rule, along with comment on the phenomenological implications of our results.

\subsection{Higher Twist in $xF_3(x,Q^2)$ Structure Functions}

In order to estimate the higher twist contribution to the $xF_3(x,Q^2)$ structure function, we have performed an analysis based on a simple model. Here the first higher twist term is extracted and to do so we have parameterised the non-singlet structure functions as

 \begin{eqnarray}
xF_3^{data}(x_i,Q^2) = xF_3^{LT}(x_i,Q^2) +\frac{h_1(x_i)}{Q^2}.
\label{EQ:9CHF3}
 \end{eqnarray}

\noindent Here leading twist(LT) term corresponds to the pQCD contribution to structure functions and the constants $h_1(x_i)$ (one per $x$ - bin) parameterize the $x$ dependence of higher twist contributions. For the leading twist term, we have utilised the results for the non-singlet structure functions obtained in our previous paper \cite{injp2}. Incorporating our results for non-singlet structure functions in NNLO as the LT terms we have extracted the difference, $xF_3^{data}(x_i,Q^2)-xF_3^{LT}(x_i,Q^2)$ from their corresponding experimental data and then fitted with $h_1(x_i)/Q^2$. From the best fitting values, we have determined the higher twist contribution terms $h_i$ per $x$-bin. In this analysis we have performed our fitting analysis within the kinematical region $0.0125 \leq x \leq 0.5$ and $1\leq Q^2 \leq 20 GeV^2$.

\ Incorporating the NNLO result (\ref{sfnnlo}) as the LT term, we have fitted the parametrization (\ref{EQ:9CHF3}) with the CCFR, NuTeV, CHORUS and CDHSW data for the $x$-bins $x_i = 0.0125, 0.015, 0.0175, 0.025, 0.035, 0.045 $. Best fitted values of $h_1$ at different values of $x$  are presented in Table \ref{TAB:9CHHTF3NS} along with the $\frac{\chi^2} {d.o.f.}$ value.

\begin{table}[h!]
\centering
\begin{tabular}{| l | l |}
  \hline
  $x_i$ & $h_1^{NNLO}$ \\
\hline
 $0.0125$ & $0.064\pm 0.0258$ \\
 \hline
  $0.015$ & $0.00504\pm 0.00804$ \\
 \hline
 $0.0175$ & $0.0189\pm 0.034$ \\
 \hline
 $0.025$ & $0.00797\pm 0.0368$ \\
 \hline
  $0.035$ & $-0.0118\pm 0.0295$ \\
 \hline
 $0.045$ & $-0.0429\pm 0.0306$ \\
 \hline
 $\frac{\chi^2}{d.o.f.}$ & 1.03 \\
 \hline
\end{tabular}
\caption{Higher Twist corrections to $xF_3(x,Q^2)$ structure function at NNLO.}
\label{TAB:9CHHTF3NS}
\end{table}

\begin{figure*}
\centering
\includegraphics[scale=0.5]{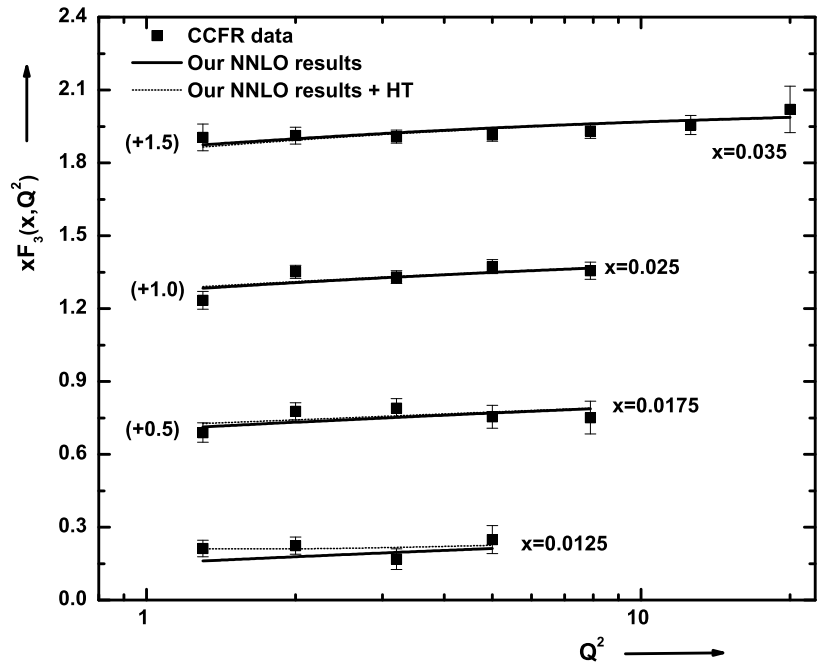}
\caption{Higher twist corrections to $xF_3(x,Q^2)$ structure function at NNLO. ($Q^2$'s are taken in the unit of $GeV^2$).}
\label{FIG:9CHHTF3NS}
\end{figure*}

\ In Fig.~\ref{FIG:9CHHTF3NS} we have presented the best fitting results of (\ref{EQ:9CHF3}) for $xF_3(x,Q^2)$ in comparison with CCFR experimental data. Here both the NNLO results, with HT and without HT are shown. Significant higher twist contribution to $xF_3(x,Q^2)$ structure function is observed in the low-$x$, low-$Q^2$ region. We observe that our expressions along with the HT corrections provide better description of CCFR data than without HT within our kinematical region of consideration.

\subsection{Higher Twist Effect in Gross-Llewellyn Smith Sum Rule}

In the previous subsection, the higher twist effects in $xF_3(x,Q^2)$ structure function is estimated by means of a simple model. We now extend the similar formalism in order to extract the higher twist contribution to the GLS sum rule associated with the $xF_3(x,Q^2)$ structure function. Here we have parameterized the sum rule as

\begin{eqnarray}
S_{GLS}^{data}(Q^2)= S_{GLS}^{pQCD}(Q^2)\bigg|_{LT}+ \frac{\mu_4}{Q^2}.
\label{EQ:9CHGLSSR}
 \end{eqnarray}

\begin{figure*}
\centering
\includegraphics[scale=0.5]{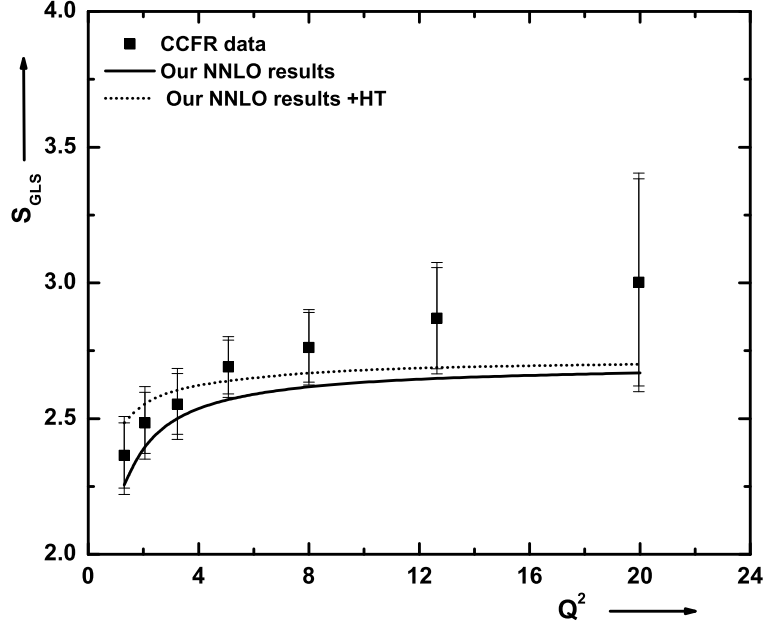}
\caption{Hgher Twist corrections to GLS sum rule at NNLO. ($Q^2$'s are taken in the unit of $GeV^2$).}
\label{FIG:9CHHTGLS}
\end{figure*}

\noindent Here leading twist(LT) term corresponds to the pQCD contribution to the GLS sum rule and $\mu_4$ signifies the contribution from first higher twist term. Incorporating the results in accord with our NNLO prediction, (\ref{srnnlo}) in (\ref{EQ:9CHGLSSR}), we have fitted the the expression with the available CCFR experimental data for GLSSR. The corresponding value of $\mu_4$ for which best fitting is obtained in NNLO are summarised in Table \ref{TAB:9CHHTGLS}  along with the respective $\frac{\chi^2} {d.o.f}$ values and in Fig.\ref{FIG:9CHHTGLS} the results for GLS sum rule with and without HT is depicted. We observe that our expressions along with the HT corrections provide better description of experimental data for GLS sum rule.

 \begin{table}[!ht]
\centering
\begin{tabular}{| l | l |}
  \hline
  & NNLO \\
\hline
 $\mu_4$ & $0.1840\pm 0.0842$ \\
 \hline
 $\frac{\chi^2}{d.o.f.}$ &  0.56 \\
 \hline
\end{tabular}
\caption{Higher Twist corrections to GLS sum rule at NNLO.}
\label{TAB:9CHHTGLS}
\end{table}

\section{Conclusion}

In this paper we present an analysis of the $xF_3(x,Q^2)$ structure function and GLS sum rule taking into account the nuclear effect and higher twist effect. In this regard, special attention is given to the nuclear shadowing effect as we are mostly concerning with the small-$x$ region. Incorporating the results of corrections due to shadowing nuclear effect obtained in earlier analysis for $xF_3(x,Q^2)$ structure function as well as GLS sum rule to our results of the structure functions and sum rules for free nucleon, we obtain structure functions and sum rules for nuclei. Nuclear correction incorporated results are studied phenomenologically and it is observed that along with the nuclear correction, our NNLO results of the $xF_3(x,Q^2)$ structure functions and GLS sum rule have the capability of providing well description of their respective experimental data collected using nuclear target. In addition, we have extracted the higher twist contributions to both the $xF_3(x,Q^2)$ structure functions and GLS sum rule using a simple model and it is observed that our NNLO expressions for $xF_3(x,Q^2)$ structure function and GLS sum rules along with the higher twist corrections provide well description of their respective experimental data.

\newpage

\end{document}